\documentclass[aps,pre,twocolumn,showpacs,floatfix,amssymb]{revtex4}
\usepackage{epsfig}

\newcommand{\ie}{{\it i.e.\ }}
\newcommand{\ignore}[1]{}

\renewcommand{\d}{{\rm d}}

\begin{document}
\title{Stable and unstable attractors in Boolean networks}
\author{Konstantin Klemm}
\affiliation{Bioinformatics Group, Dept.\ of Computer Science,
University of Leipzig, Kreuzstr.~7b, D-04103 Leipzig, Germany}
\author{Stefan Bornholdt}
\affiliation{Institute for Theoretical Physics,
University of Bremen, Otto-Hahn-Allee, D-28359 Bremen, Germany}
\date{\today}

\begin{abstract} 
Boolean networks at the critical point have been a matter of 
debate for many years as, e.g., scaling of number of attractor 
with system size. Recently it was found that this number scales 
superpolynomially with system size, contrary to a common earlier 
expectation of sublinear scaling. We here point to the fact that 
these results are obtained 
using deterministic parallel update, where a large fraction of 
attractors in fact are an artifact of the updating scheme. 
This limits the significance of these results for biological 
systems where noise is omnipresent. We here take a fresh look at 
attractors in Boolean networks with the original motivation 
of simplified models for biological systems in mind. We test  
stability of attractors w.r.t.\ infinitesimal deviations from 
synchronous update and find that most attractors found under parallel 
update are artifacts arising from the synchronous clocking mode. 
The remaining fraction of attractors are stable against fluctuating 
response delays. For this subset of stable attractors we observe 
sublinear scaling of the number of attractors with system size. 
\end{abstract} 

\maketitle
Boolean networks at the critical point (sometimes also called  Kauffman
networks) have been discussed as simplified models for 
gene regulation networks for many years  
\cite{Kauffman69,Glass73,Kauffman93}.  We currently experience a
resurgence of interest in these models,  as structure and dynamics of the
genetic network in a living cell  become visible thanks to new powerful
experimental techniques  (DNA chips) \cite{NatureDNAchips}. From the
theorist's point of view, Boolean networks exhibit interesting statistical
mechanics with a prominent order/disorder phase transition
\cite{BooleanReview}. Earlier, the critical state has been postulated to have
some relevance in the biological  context as the scaling properties of numbers
of attractors with  network size appeared to resemble how number of cell types
scale  with amount of genetic information when comparing simple and complex 
organisms \cite{Kauffman93}. Until recently it was believed that the total
number of attractors increased as $\sqrt{N}$ \cite{Kauffman93}. This has been
falsified by improved simulation  techniques \cite{Bilke01} and it was shown
that the  total number of attractors grows faster than any polynomial 
\cite{Samuelsson03,Drossel04}.

Let us here step back for a moment and reconsider Kauffman networks 
in the context of their original motivation, as models for biological 
systems. While the use of models discrete in time is an established 
approach in many circumstances of biological modeling, such idealizations 
always have to be treated with special care. In the case of Kauffman networks, 
the system evolves by a synchronous update of all nodes at integer values of 
time. Such a clocking, however, can produce spurious synchrony. For instance,
subsystems are kept phase synchronized even if they are not interacting at all.
In order to circumvent computational artifacts it has been suggested to use
a more natural updating schedule \cite{Huberman93}. For example it has been 
shown that the complex spatio-temporal patterns observed under synchronous
update often disappear when units are updated asynchronously 
\cite{Choi83,Ingerson84,Huberman93,Bagley96,Harvey97,Klemm03}.
 
\begin{figure}[hbt]
\centerline{\epsfig{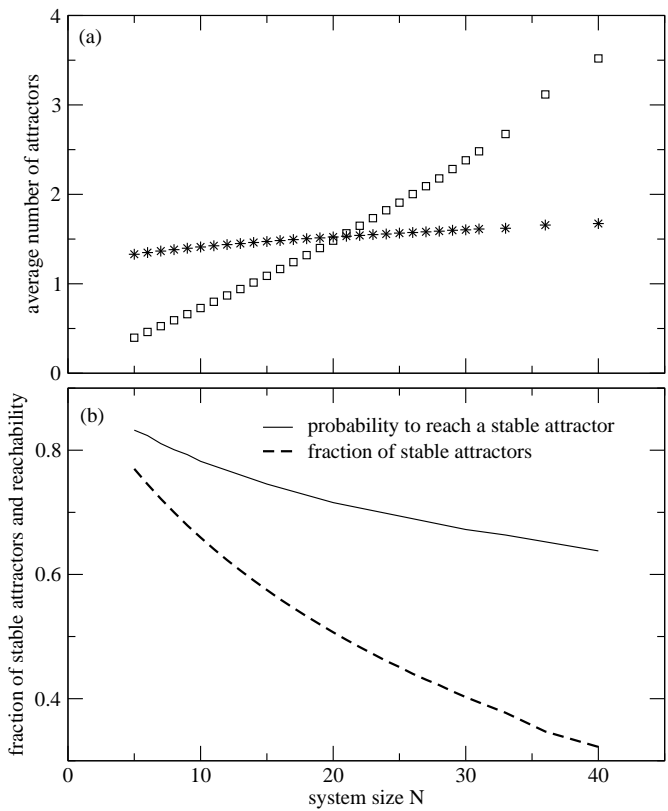}}
\caption{\label{fig:stabf_0} Frequency and accessibility of stable and
unstable cycles in Kauffman networks. (a) Average number of stable 
($\ast$) and unstable ($\Box$) attractors as a function of the number of 
nodes in the network. 
(b) Fraction of initial conditions leading to a stable attractor
(solid line) and the ratio between numbers of stable and all attractors
(dashed line). Data points in (a) are averages over $R$ independent 
realizations of the Kauffman network, $R=10^7$ for $N<24$, $R=10^6$ for 
$24\le N \le31$ and $R=10^4$ otherwise. For increasing computational 
speed the networks are subject to the decimation procedure \cite{Bilke01} 
before simulation. For the decimated network we fully enumerate the state 
space such that it is certain that all cycles are detected. Sizes of 
basins of attraction in (b) have been estimated in $10^5$ networks,
testing 100 randomly chosen initial conditions in each original 
network (no decimation applied).}
\end{figure}

In this paper we reconsider Boolean networks at criticality, while 
destroying spurious synchrony by equipping the nodes with weakly 
fluctuating response delays. This allows us to analyze the stability 
of dynamical attractors in the discrete network model. A deterministic 
Kauffman network at an attractor is perturbed by a slight shift of
update events forward or backward in time. If all such perturbations die
out, i.e.\ the system returns to the identical attractor, we call this 
attractor ``stable''. Otherwise ongoing temporal fluctuations
accumulate and drive the system away from the attractor. These latter 
cases correspond to attractors that are an artifact caused by 
synchronous update of the deterministic system. When systematically 
applying this method to Kauffman networks we obtain as main result 
that the number of stable attractors grows sublinearly with system size
(see Fig.\ \ref{fig:stabf_0}(a)). 

Let us study a Kauffman network composed of $N$ binary nodes where 
each node determines its state $x_i$ by applying a Boolean function 
(a rule table) $f_i:\{0,1\}^2 \rightarrow\{0,1\}$ on inputs received 
from two other nodes $a(i)$ and $b(i)$, according to a randomly 
chosen quenched topology. To be definite, we exclude self-couplings. 
Starting from an arbitrary initial
condition $(x_1(0),x_2(0),\dots,x_N(0))$, states of the nodes are 
synchronously updated at integer times $t$ according to the Boolean 
function  
\begin{equation} \label{eq:booldyn}
x_i(t+1) = f_i(x_{a(i)}(t),x_{b(i)}(t)). 
\end{equation}
The network itself as defined by $f_i$, $a(i)$, and $b(i)$ remains 
constant in time. Running the system from a randomly chosen initial 
condition, its finite discrete state space of $2^N$ possible states 
ensures that, eventually, a state reappears that has been encountered 
before. From thereon, the deterministic system will indefinitely 
follow the attractor it reached (which is either a periodic limit 
cycle or a fixed point). Different initial conditions may lead to 
the same or to a different attractor. The total number of attractors
is a characteristic property of a network. The expected number of 
cycles in an ensemble of random Kauffman networks has been shown 
to increase superpolynomially with system size $N$ \cite{Samuelsson03}.

Let us now define a criterion for stability of an attractor 
in the presence of deviations from deterministic parallel update. 
For this purpose, we replace the discrete update times by a continuous 
time variable $t$ where nodes may update at any point in time. 
Our goal is to slightly desynchronize the dynamics of the network 
by shifting the individual updates of nodes to slightly earlier 
or later time points. To avoid that this generates spurious 
spikes during transitional phases (e.g.\ when several signals 
interact that used to be simultaneous, but now arrive at 
slightly different times), nodes have to be prevented from switching 
on a time scale $s$ much shorter than the original integer update
time step (i.e., $s \ll 1$). This is implemented by a low pass filter 
that averages out fluctuations on short characteristic times 
scales $s$, namely by averaging over the input signal according to
\begin{equation} \label{eq:booldynf}
x_i(t+1) = \Theta \left[ (2s)^{-1} \int_{t-s}^{t+s}
f_i(x_{a(i)}(t),x_{b(i)}(t))\d t - 1/2\right]
\end{equation}
where $\Theta$ is the step function with $\Theta(h)=1$ for $h\geq 0$ 
and $\Theta(h)=0$ otherwise. Let us briefly check how this works. 
Imagine node $a(i)$ switches on at time $t$, node $b(i)$ switches 
off at time $t+\epsilon$, while $f_i$ is the function {\sc and}. 
Without low pass filter ($s=0$), node $i$ switches on at time $t+1$ 
and off again at time $t+1+\epsilon$. When the switching time scale 
$s$ of nodes is sufficiently slowed down, $s>\epsilon$, the spurious 
spike is filtered out, i.e.\ node $i$ remains constant. 
Note that in the limit of fast switching time scale 
$s\rightarrow 0$, Eq.\ (\ref{eq:booldynf}) converges towards Eq.\ 
(\ref{eq:booldyn}). In particular, all synchronous solutions of 
Eq.\ (\ref{eq:booldyn}), i.e.\ solutions with nodes switching 
precisely at integer values of $t$, are solutions of Eq.\ 
(\ref{eq:booldynf}) for arbitrary $s<1/2$, as well. 

Starting from such a synchronous attractor of a network, let us 
now perturb it at some time $T$ by temporarily retarding parts 
of the switching events.
Thereby, a subset of nodes that would normally change state at 
time $T$ is kept frozen in their present state during the time 
interval $[T,T+\epsilon]$ with $\epsilon<s$. After $t = T+\epsilon$, 
we let the system evolve as usual according to Eq.\ (\ref{eq:booldynf}). 
Note that the original and the perturbed solution
differ only on time intervals $[t,t+\epsilon[$ for integer $t$. 
In general, the time lag may propagate, i.e.\ for each integer $t\ge T$
some units flip at time $t$ while others flip at time $t+\epsilon$ in the
perturbed solution. If, however, there is a later integer time $t^\ast>T$ 
such that either no flips occur at time $t^\ast$ or no flips happen at
time $t^\ast+\epsilon$, the perturbation has been overcome and the system 
has regained synchrony. 
We call an attractor stable if for all possible perturbations of the above type
(i.e.\ for all possible permutations of perturbed and non-perturbed nodes) 
the system regains synchrony and the original attractor is stabilized within 
a finite time interval.
Otherwise the attractor is called unstable. 
In real world situations with continuous noise, such unstable 
attractors will accumulate phase shifts that eventually shift the 
system into some other, stable attractor. 

With this we here choose a particularly simple criterion for the stability of 
an attractor in a Boolean network. The system is on a stable attractor if 
after each small perturbation it reaches the attractor again where, 
as a minimal perturbation, a small deceleration or acceleration of 
a switching event is used. On unstable attractors, such time lags do not 
relax. Thus, ongoing perturbations eventually lead to a change in time ordering 
of the switching events and the system reaches a different attractor.
Note that this scenario is much better suited as a stability criterion than 
stochastically adding or removing switching events \cite{Qu02,Amaral04}, which 
does not allow for the limit of infinitesimally weak perturbations. 
The low pass filter characteristics used here is further motivated by 
the dynamics of biochemical switches \cite{Rao02} where 
molecule concentrations typically respond slowly, leading to an 
overall low-pass filter characteristics of the switch. 
Low pass filter characteristics is a natural property of genes and is 
a simple means of stabilization which is of low cost and ubiquitous in nature. 
We here model this mechanism as an effective time average over the input 
signal that suppresses short pulses. Note that under synchronous update 
of the model networks, pulses cannot be shorter than unit time such that 
the filter can be neglected.

Applying the robustness criterion to random Boolean networks at criticality, 
one observes that the average number of all attractors, stable and 
unstable ones, grows much faster than the average number of stable
attractors alone (see Fig.\ \ref{fig:stabf_0}(a)). In large networks, 
almost all attractors are expected to be unstable. Interestingly, 
the probability to reach a stable attractor from a random initial 
condition is much larger than the fraction of stable attractors,  
as shown in Fig.\ \ref{fig:stabf_0}(b). Thus, unstable attractors 
typically have significantly smaller basins of attraction than 
stable attractors.
\begin{figure}[hbt]
\centerline{\epsfig{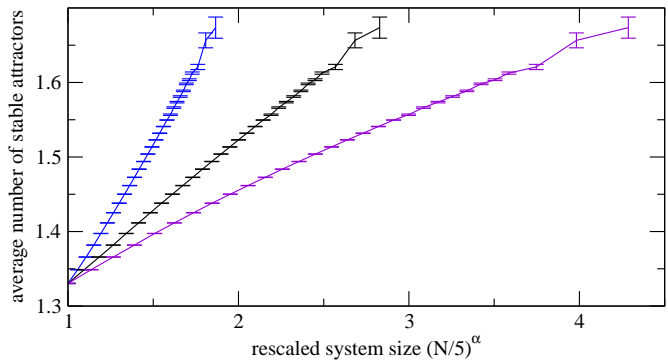}}
\caption{\label{fig:stabf_1} System size scaling of the number of 
stable attractors. Plotted as a function of the rescaled number of 
nodes $(N/5)^\alpha$ with $\alpha = 0.3, 0.5, 0.7$ (left to right).
Error bars in (b) indicate standard deviation divided by $\sqrt{R}$ 
with ensemble size $R$ (cf.\ Fig.\ \ref{fig:stabf_0}).}
\end{figure}
The main result is that, with system size $N$, the number of stable 
attractors grows sublinearly as $\sim N^\alpha$ with $\alpha \approx 0.5$,
as shown in Fig.\ \ref{fig:stabf_1}.
A least squares fit of the form $c + bN^\alpha$ fits best ($\langle \chi^2
\rangle = 0.00013$) with the parameter values $\alpha=0.448$, $c = 1.107$, 
and $b = 0.108$ (with a correlation coefficient for this fit of $r=0.999742$).
\begin{figure}[hbt]
\centerline{\epsfig{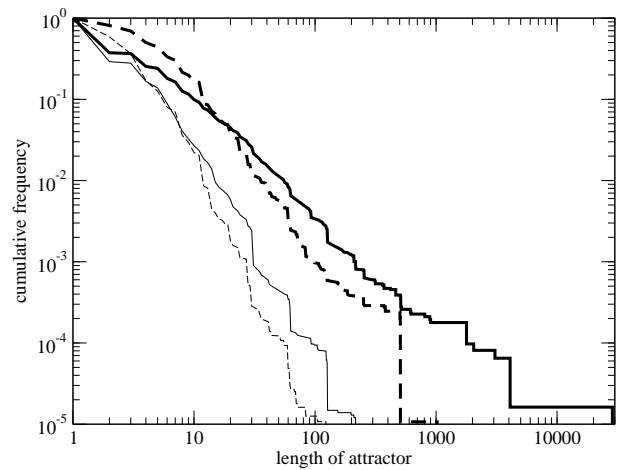}}
\caption{\label{fig:lengths_0} Statistics of attractor lengths for
networks with $N=10$ (thin curves) and $N=30$ (thick curves).
The cumulative distribution is shown for stable attractors (solid
lines) and unstable attractors (dashed lines).
For $N=10$ nodes the average length of stable attractors is
$\langle l \rangle_{\rm nat} = 2.59$, of unstable attractors 
$\langle l \rangle_{\rm art} = 3.56$;
for $N=30$ we find
$\langle l \rangle_{\rm nat} = 5.50$ and
$\langle l \rangle_{\rm art} = 7.16$.
}
\end{figure}
Further let us analyze the number $l$ of states contained in the attractors.
While stable attractors are shorter on average than unstable attractors,
the distribution of attractor lengths is broader for stable 
than for unstable attractors, as shown in Fig.\ \ref{fig:lengths_0}. 
The majority of long attractors with lengths far above average are stable.


\begin{figure}[hbt]
\centerline{\epsfig{file=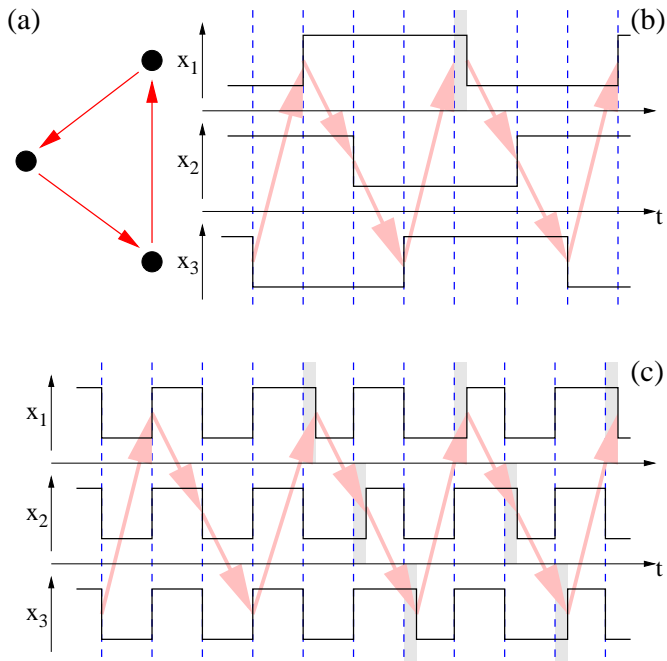,width=.49\textwidth}}
\caption{\label{fig:3loop_illu} 
A stable and an unstable attractor in the same system.
(a) Three nodes forming a directed cycle (feedback loop).
Each node $i$ performs the Boolean function negation on the input from its
predecessor $j$, \ie $x_i(t+1) = 1-x_j(t)$.
(b) Temporal evolution of a stable attractor. There is a unique causal chain
of flipping events (thick arrows).
A retarded update (shaded box) retards all subsequent events by the same
amount of time. Thus perturbations heal immediately.  
(c) Unstable attractor. It can be interpreted as the superposition of
three independent chains of flipping events propagating along the cycle
of nodes. One of the chains is indicated by thick arrows as in (b).
Retarding an event effects subsequent events in the same causal chain only.
Causal chains remain phase shifted.
The system does not regain synchrony after a perturbation.
See reference \protect \cite{Klemm04} for a detailed analysis
of attractor stability in small systems.
}
\end{figure}
Can we understand by a simple picture how unstable attractors differ from stable
ones?
Most obviously, unstable attractors occur when the network falls into two or more
non-interacting clusters. When all updates in one of the clusters are delayed
by the time $\epsilon$, this phase lag with respect to all other clusters
cannot heal. All attractors with flipping events in more than one network
cluster are unstable. However, also in networks consisting of a single cluster (more
precise: with a single strongly connected component) unstable
attractors are found. Figure \ref{fig:3loop_illu} illustrates the coexistence
of a stable and an unstable attractor in a small connected network. The
example suggests that an attractor is stable if there is a single cascade of
switching events. Let us consider 
 the minimal number of simultaneous flipping events 
\begin{equation}
m = \min_t |\{ i | x_i(t)\neq x_i(t+1)\} |
\end{equation}
for a given attractor. The attractors with $m=0$ are the fixed points. These are
stable by definition because no flipping events are to be retarded. Attractors
with $m=1$ are stable as well. These attractors contain a step with only a single
flipping event. Going through this step the system always regains synchrony.
For $m\ge 2$ the attractor is likely to contain several chains of causal events as in
Fig.\ \ref{fig:3loop_illu}(c). In the simulations we find that a large fraction
$u$ of the attractors with $m\ge 2$ is unstable,
$u=0.856, 0.882, 0.899, 0.9094$ for $N=10,20,30,36$, respectively.
Thus the minimal number of simultaneous flipping events $m$ allows for an almost
perfect distinction between stable and unstable attractors. Note that $m$
is measured in the decimated networks \cite{Bilke01}.

Comparing these results to past studies of random Boolean networks 
at criticality ($K=2$ inputs per node), we obtain a distinctly different 
picture: Only a small fraction of all attractors are at all stable 
against small amounts of noise. Or, put differently, the effect of 
spurious synchronization due to a parallel update mode has been 
underestimated in previous studies, at least where these studies have 
been made with a potential application to biological systems in mind. 
In particular, characteristic properties of the attractor statistics 
are different when considering the subset of stable attractors:  
The average number of stable attractors scales less than linearly 
while the number of unstable attractors shows a faster, superlinear growth
with $N$. Also, stable attractors have a significantly larger basin of 
attraction than unstable ones. One may speculate that this latter 
property might have been the reason for the long prevalence of the 
opinion that the total number of attractors scales as $\sqrt{N}$ 
\cite{Kauffman93}.  Mainly these stable attractors were likely to be found
in the early studies using sparse sampling. 

If one aims at discussing Boolean networks as simple models for 
biological systems, our study suggests to consider more carefully the 
question of which attractors are at all relevant to the biological system. 
For example, Kauffman's observation of the number of attractors 
in critical random Boolean networks exhibiting a similar scaling with 
system size as the number of cell types with genome size in 
organisms \cite{Kauffman93} seems to be wrong in the light of the results by 
Troein and Samuelsson \cite{Samuelsson03}. However, it appears to be 
still open to debate when considering solely the subset of stable attractors. 
\medskip
 
{\bf Acknowledgement} We thank Barbara Drossel, Albert Diaz-Guilera, 
Leon Glass, and Bj\" orn Samuelsson for helpful comments. 
This work was supported by the Deutsche Forschungsgemeinschaft DFG.


\medskip

\begin{thebibliography}{99}

\bibitem{Kauffman69}
S.A.\ Kauffman, J.\ Theor.\ Biol.\ {\bf 22}, 437 (1969).  

\bibitem{Glass73} 
L.\ Glass and S.A.\ Kauffman, J.\ Theor.\ Biol.\ {\bf 39}, 103 (1973).

\bibitem{Kauffman93}
S.A.\ Kauffman, {\em The Origins of Order.} (Oxford University Press, New York, 1993)

\bibitem{NatureDNAchips}    
For a review see for example {\em The chipping forecast II}, Nature Genetics Supplement {\bf 32}, 461-552 (2002).

\bibitem{BooleanReview}
M.\ Aldana, S.\ Coppersmith and L.P.\ Kadanoff
in {\em Perspectives and Problems in Nonlinear Science},
edited by E.\ Kaplan, J.E.\ Marsden, and K.R.\ Sreenivasan (Springer, New York, 2003).

\bibitem{Bilke01} 
S. Bilke and F. Sjunnesson, Phys.\ Rev.\ E {\bf 65}, 016129 (2001).

\bibitem{Samuelsson03}
B.\ Samuelsson and C.\ Troein, Phys.\ Rev.\ Lett.\ {\bf 90}, 098701 (2003).

\bibitem{Drossel04}
B.\ Drossel, T.\ Mihaljev, F.\ Greil, preprint cond-mat/0410579

\bibitem{Huberman93}
B. A. Huberman, N. S. Glance,
Proc.\ Natl.\ Acad.\ Sci.\ USA {\bf 90}, 7716 (1993).

\bibitem{Choi83}
M.Y.\ Choi and B.A.\ Huberman, Phys.\ Rev.\ B {\bf 28}, 2547 (1983)

\bibitem{Ingerson84}
T.E.\ Ingerson and R.L.\ Buvel, Physica D {\bf 10}, 59 (1984).

\bibitem{Bagley96}
R.J.\ Bagley   and L.\ Glass, J.\ Theor.\ Biol.\ {\bf 183} 269 (1996). 

\bibitem{Harvey97}
I.\ Harvey and T.\ Bossomaier, {\em Proceedings of the Fourth European Conference 
on Artificial Life (ECAL97)}, pp.\ 67-75 (MIT Press, 1997).

\bibitem{Klemm03}
K.\ Klemm, S.\ Bornholdt, preprint q-bio/0309013 (2003). 

\bibitem{Qu02}
X.\ Qu, M.\ Aldana and L.P.\ Kadanoff, J.\ Stat.\ Phys.\ {\bf 109} 967, (2002).

\bibitem{Amaral04}
L. A. N. Amaral, A. D\'\i az-Guilera, A. A. Moreira, A. L. Goldberger, and
L. A. Lipsitz,
Proc.\ Natl.\ Acad.\ Sci.\ USA {\bf 101}, 15551 (2004).

\bibitem{Rao02}
C.V.\ Rao, D.M.\ Wolf, and A.P.\ Arkin, Nature {\bf 420}, 231-237 (2002). 

%
\bibitem{Klemm04}
K.\ Klemm, S.\ Bornholdt, preprint q-bio/0409022 (2003).
%

\end{thebibliography}
\end{document}